# On How Developers Test Open Source Software Systems

Andy Zaidman, Bart Van Rompaey, Serge Demeyer, Arie van Deursen

# On How Developers Test Open Source Software Systems

Andy Zaidman[1], Bart Van Rompaey[2], Serge Demeyer[2], and Arie van Deursen[3]

[1]Delft University of Technology, The Netherlands – a.e.zaidman@tudelft.nl
[2]University of Antwerp, Belgium – {bart.vanrompaey2, serge.demeyer}@ua.ac.be
[3]Delft University of Technology & CWI, The Netherlands – arie.vandeursen@tudelft.nl

**Abstract**

*Engineering software systems is a multidisciplinary activity, whereby a number of artifacts must be created — and maintained — synchronously. In this paper we investigate whether production code and the accompanying tests co-evolve by exploring a project's versioning system, code coverage reports and size-metrics. Three open source case studies teach us that testing activities usually start later on during the lifetime and are more "phased", although we did not observe increasing testing activity before releases. Furthermore, we note large differences in the levels of test coverage given the proportion of test code.*

## 1  Introduction

Lehman has taught us that a software system must evolve, or it becomes progressively less useful [18, 19]. For many people evolving a software system has become a synonym for adapting the source code as this concept stands central when thinking of software. Software, however, is multidimensional, and so is the development process behind it. This multidimensionality lies in the fact that to develop high-quality source code, other artifacts are needed, e.g. specifications, constraints, documentation, tests, etc. [20].

In this paper we explore two dimensions of the multidimensional software evolution space, as we focus on how the basic software evolves with regard to the accompanying tests of the software system. To characterize why tests are so important during evolution, we first discuss three major objectives:

*Quality assurance* Tests are typically engineered and run to ensure the quality of a software system [6, 21].

*Documentation* In Agile software development methods such as eXtreme Programming (XP), tests are explicitly used as a form of documentation [7, 8].

*Confidence* At a more psychological level, tests help the software (re)engineer become more confident, because of the safety net that is provided by the tests [5, 7].

Another aspect of testing that cannot be neglected is the impact on the economy of the software development process: testing is known to be very time-intensive. Estimates by Brooks put the total time devoted to testing at 50% of the total allocated time [3, 22], while Kung et al. suggest that 40 to 80% of the development costs of building software is spent in the testing phase [17].

Knowing the necessity of a software system's evolution, the importance of having a test suite available and the cost-implications of building (and maintaining) a test suite, we wonder how test and production code co-evolve during a software project's lifetime. Ideally, we understand that test code and production code should be developed and maintained synchronously, for at least two reasons:

- Newly added functionality should be tested as soon as possible in the development process [2].
- When changes, e.g. refactorings, are applied, the preservation of the behavior needs to be checked [7, page 159].

In this context Van Deursen et al. have shown, that even while refactorings are behavior preserving, they potentially invalidate tests [9]. Elbaum et al. came to the conclusion that even minor changes in production code can have serious consequences on test coverage, or the fraction of production code tested by the test suite [11]. These observations reinforce the claim that production code and test code need to co-evolve.

This leads to the almost paradoxical situation whereby tests are quasi essential for the success of the software (and its evolution), while also being a serious burden during maintenance. This brings us to our central question:

*How does test co-evolution happen in real world, open source systems?*

We refine this question into a number of subsidiary research questions:

**RQ1** How can we summarize the co-evolution of test code and production code?

**RQ2** Does co-evolution always happen synchronously or is it phased?

**RQ3** Can an increased testing effort be witnessed right before a major release or other event in the project's lifetime?

**RQ4** Can we detect testing strategies, e.g. test-driven development?

**RQ5** Is there a relation between test effort and test coverage?

The next section introduces three views on this two-dimensional software evolution space and we discuss the views with a running example. Sections 4 through 6 present our three case studies on respectively CheckStyle, PMD and ArgoUML. Section 7 relates test coverage to the fraction of test code written. Section 8 provides discussion, while Section 10 presents our conclusion and future work.

## 2 Co-evolution Recovery

As studying the history of software projects involves large amounts of data, we make use of visualizations to answer evolution-related questions. More specifically, we introduce three distinct views, namely:

1. The **change history view**, wherein we visualize the *commit*-behavior of the developers.
2. The **growth history view** that shows the relative growth of production code and test code over time.
3. The **coverage evolution view**, where we plot the test coverage of a system at discrete times.

To introduce these three views, we use JPacman as a running example. JPacman, a teaching example for the software testing course at the Delft University of Technology, has been developed using a test-intensive XP-style process, featuring unit and integration tests achieving a high level of test coverage. Due to its simplicity and the inside knowledge available, it perfectly fits the purpose of a running example.

### 2.1 Change History View

**Introduction.** Visualizing the revision history of a set of source code entities has been used to study how these entities co-evolve, e.g. the work of Gîrba and Ducasse [14], Van Rysselberghe and Demeyer [25] and Wu et al. [26]. Other research in the same area does not rely on visualizations but still identifies logical coupling, e.g. Gall et al. [12] and Ball et al. [1].

Typically, these visualizations use one axis to represent time, while the other represents the source code entities. This visualization-approach has been used to detect logical coupling between files, determine the stability of classes over time, etc. These approaches however, do not make a clear distinction between different types of source code entities, e.g. between production code and test code.

**Goal.** The Change History View allows us to learn whether (i) production code has an associated (unit) test and (ii) whether these are added and modified at the same time. As such, we seek to answer RQ2 and RQ4.

**Description.** In this view:
- We use an XY-chart wherein the X-axis represents time and the Y-axis source code entities.
- We make a distinction between production files and test files. Unit tests that test a specific unit are placed on the same horizontal line.
- We make a distinction between files that are introduced and files that are modified.
- We use colors to differentiate between newly added production code (red), modified production code (blue), newly added tests (green) and modified tests (yellow).

**Interpretation.** An example of our change history plot can be seen in Figure 1 which visualizes the commit behavior of JPacman's single developer[1].

We are looking for patterns in the plotted dots that signify co-evolution. Test files introduced together with the associated production unit are represented as green dots plotted on top of red dots. Test files that are changed alongside production code show as yellow dots on top of blue dots. Vertical green or yellow lines indicate many changes to the test code, whereas horizontal lines stand for frequently changed files (not visible in JPacman).

In the JPacman case, we notice that in the first version a lot of test cases are introduced alongside production units. This indicates that either (i) the project had a history before it was brought into the versioning system; or (ii) a test-driven-like approach has been used [2]. In this case, the project had a history. Other evidence of testing is present but happens at random moments outside major change periods.

**Technicalities.** The correlation between production and test code happens on the basis of file naming conventions (e.g. a test case that corresponds to a certain production class has the same file name with postfix *"Test"*). Unit tests that cannot be correlated are considered to be integration tests and are placed on the top lines of the graph. For completeness' sake, we add that the projects under consideration during this study adhere to these naming conventions quite well, with only a handful of outliers being found.

---

[1] Ideally, these visualizations should be seen in color. High-resolution color images are also available at http://swerl.tudelft.nl/testhistory

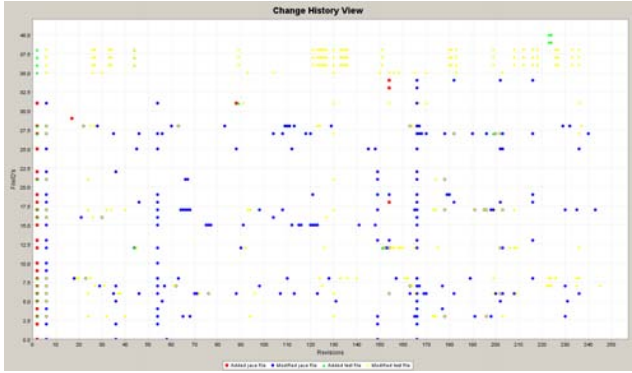

**Figure 1. Change history view of JPacman.**

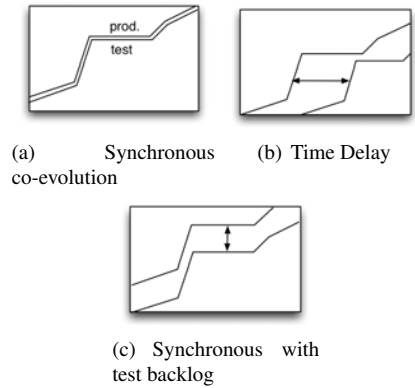

(a) Synchronous co-evolution  (b) Time Delay

(c) Synchronous with test backlog

**Figure 2. Patterns of synchronous co-evolution.**

Note that the number of units shown in this visualization is often higher than the number of classes present in the latest version of a software system. This is due to the fact that when a file gets deleted at a certain point in time, it remains present in the visualization. In this context, we also want to note the presence of *"outliers"* in the visualization (e.g., see Figure 5), dots that lie above the growing curve of classes that are added. These outliers are caused by successive *move* operations in the subversion repository, but remain associated with their original introduction date.

### 2.2 Growth History View

**Motivation.** The use of source code metrics to characterize the evolution of a system has for example been used by Godfrey and Tu to investigate whether open source software and commercial software have different growth rates [15] or by Gall et al. to identify possible shortcomings in a system [13]. To a certain degree, our research interests are similar as we investigate whether production code and test code grow at similar or different points in time during a project's history.

**Goal.** It is our aim to identify growth patterns indicating (non-)synchronous test and production code development (RQ2), increased testing effort just before a major release (RQ3) and evidence of test-driven development (RQ4).

**Description.** In this view:
- We use an XY-chart to plot the co-evolution of a number of size metrics over time.
- The five metrics that we take into consideration are: Lines of production code (pLOC), Lines of test code (tLOC), Number of production classes (pClasses), Number of test classes (tClasses) and Number of test commands[2] (tCommands).
- In addition to these five metrics, we also visualize two derived metrics, namely:

$$pClassRatio = pClass/(tClass + pClass) \times 100$$

---
[2] A test command is a container for a single test [24].

$$pLOCRatio = pLOC/(tLOC + pLOC) \times 100$$

- Metrics are presented as a cumulative percentage chart up to the last considered version, as we are particularly interested in the co-evolution and not so much in the absolute growth.
- The X-axis is annotated with release points.

**Interpretation.** First of all, we can observe phases of relatively weaker or stronger growth throughout a system's history. Typically, in iterative software development new functionality is added during a certain period after a major release, after which a "feature freeze" [16] comes into play allowing no more new functionality to be added. At that point, bugs get fixed, testing effort is increased and documentation written.

Secondly, the view allows us to study growth co-evolution. We observe (lack of) synchronization by studying how the measurements do or do not evolve together in a similar direction. Production and test effort is spent synchronously when the two curves are similar in shape. A horizontal translation indicates a time delay between one activity and a related one, whereas a vertical translation signifies that a historical backlog has been accumulated over time for one activity compared with another. Figure 2 presents a schematic example of three typical co-evolution situations. Note that in Figures 2(a) & 2(b), when determining the backlog or time delay, the baseline situation is the last considered version. At that point, both curves reach 100%, indicating that the effort of writing production and test code is in balance.

Thirdly, the interaction between measurements yields valuable information as well. We now refer to Table 1, in which a number of these interactions are outlined. Take for example the first line in Table 1, where it is indicated that an increase in production code and a constant level of test code (with the other metrics being unspecified) points towards a "pure development" phase.

In the case of JPacman (Figure 3), we notice that 90% of production code was introduced during the first few commits. Therefore, we can conclude that the system had been growing before being entered into the versioning system, as already mentioned in the Change History View. Overall, we can distinguish three phases in the history. Until about release 1.8, production code has been more or less status quo (with certain temporal reductions in size), while the amount of testing code has been steadily increasing. Around release 1.8 production code stabilizes, but all test entities drop, indicating that tests are being restructured. In the last period, development happens more synchronously (annotation 1) with somewhat more emphasis on testing to reduce the historical testing backlog. At around version 1.2 and 1.3, we observe periods of pure testing (ann. 2 and 3). These observations are backed up by inspections of the log messages.

| pLOC | tLOC | pClasses | tClasses | tCommands | interpretation |
|---|---|---|---|---|---|
| ↗ | → |  |  |  | pure development |
| → | ↗ |  |  |  | pure testing |
| ↗ | ↗ |  |  |  | co-evolution |
| ↗ | ↗ |  | → | → | test refinement |
| → | → | ↗ | ↗ |  | skeleton co-evolution |
|  | → |  | ↗ |  | test case skeletons |
|  | → |  |  | ↗ | test command skeletons |
| → | ↘ |  |  |  | test refactoring |

**Table 1. Co-evolution scenarios.**

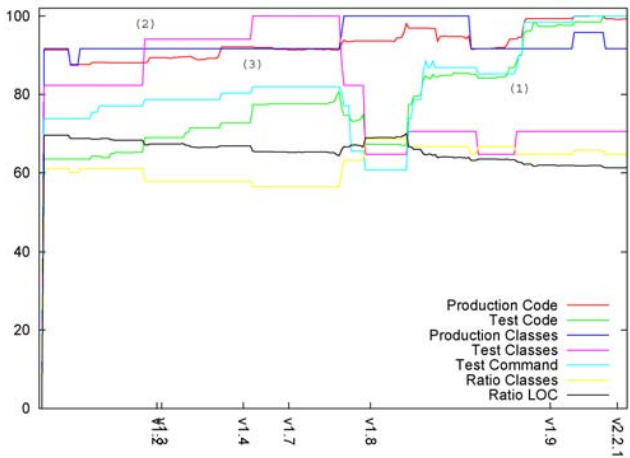

**Figure 3. Growth History view of JPacman.**

**Technicalities.** To separate production classes from test classes we use regular expressions to detect whether a class is a jUnit test case. As a first check, we look at whether the class extends `junit.framework.TestCase`. If this fails, e.g. because of an indirect generic test case [24], we search for a combination of `org.junit.*` imports and `setUp()` methods.a basic block is a sequence of bytecode instructions without any jumps or jump targets

Counting the number of test commands was done on the basis of naming conventions. More specifically, when we found a class to be a test case, we looked for methods that would start with `test`. We are aware that with the introduction of jUnit 4.0, this naming convention is no longer necessary, but the projects we considered still adhere to them.

### 2.3 Coverage Evolution View

**Motivation.** Test coverage is often seen as an indicator of "test quality" [27]. Therefor, our third view represents the coverage of a system over time, providing not only a quality-driven view, but also a *health*-driven view, representing long-term quality.

**Goal.** To be able to judge the long-term *"test health"* of a software project.

**Description.** In this view:
- We use an XY-chart representing time (in terms of releases) on the X-axis and the overall test coverage percentage on the Y-axis.
- We plot four coverage measures: class, method, statement and block[3] coverage.

**Interpretation.** Constant or growing levels of coverage over time indicate good testing health, as the testing-process is under control. Severe fluctuations or downward spirals imply weaker test health. In Figure 4 we present JPacman's coverage measurements. The coverage remains fairly constant, but around release 1.8 a drop can be witnessed, which coincides with the migration of jUnit 3.8 to 4.0 and a cleanup of the test-code.

**Technicalities.** For now we only compute the test coverage for the major and minor releases of a software system and are thus not computing coverage for every commit as: (i) we are specifically interested in long-term trends, (ii) computing test coverage (for a single release) is time-consuming and (iii) automating this step for all releases proved difficult, due to changing build systems and (varying) external dependencies that were not always available in the version management system.

## 3 Experimental setup

**Toolchain** Our toolchain[4] is built around the *Subversion*[5] version management system. With the help of the *cvs2svn*[6] script we are also able to deal with *CVS*. Using Subversion and the *SVNKit* library[7], we are able to query the subversion repository directly from our Java-built toolchain that automatically generates the change history view (Section 2.1) and the growth history view (Section 2.2).

---
[3] A basic block is a sequence of bytecode instructions without any jumps or jump targets, also see http://emma.sourceforge.net/faq.html (accessed April 13, 2007)
[4] Download from: http://swerl.tudelft.nl/testhistory
[5] http://subversion.tigris.org/
[6] http://cvs2svn.tigris.org/

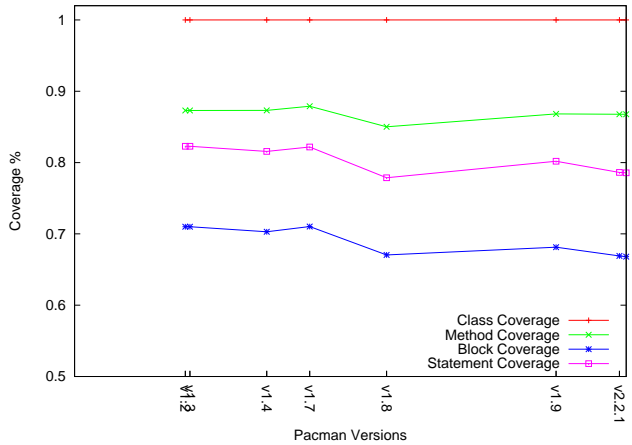

**Figure 4. Coverage evolution view of JPacman.**

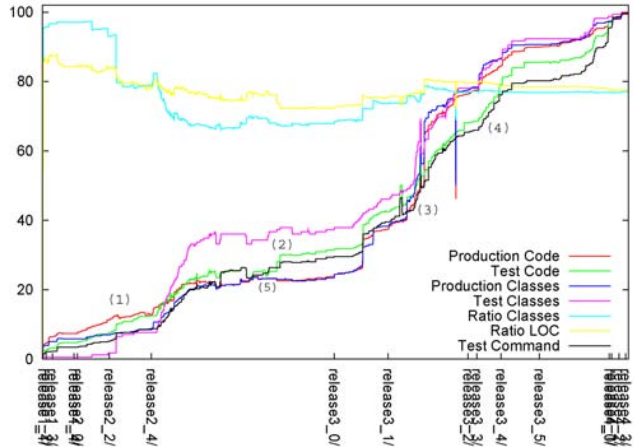

**Figure 6. Checkstyle growth history view.**

For the coverage history view, we used *Emma*[8], an open source test coverage measurement solution. We integrated Emma in the Ant build process of the case studies with the help of scripts and manual tweaking, as automating this process proved difficult.

**Case studies** As case studies, we selected Checkstyle, PMD and ArgoUML. Our main criteria for selecting the case studies were: (i) the possibility of having a local copy of the CVS or Subversion repository, for performance reasons, (ii) Java, as our toolchain is targeted towards Java, and (iii) the availability of jUnit tests.

When discussing the case studies, note that not every type of visualization is shown for each case study, due to space restrictions. However, all views can be seen in the online appendix[9].

## 4 Case 1: Checkstyle

**Introduction.** Checkstyle[10] is a tool that checks whether Java code adheres to a certain coding standard. For Checkstyle, six developers made 2260 commits in the interval between June 2001 and March 2007, resulting in 738 classes and 47KLOC.

**Change history view.** The change history view of Checkstyle (Figure 5) shows the addition of production code files by date. Figure 5 resulted in the following observations with regard to the testing behavior of the developers. At the very beginning of the project up until commit #280, there is only one test. At that point, a number of new tests are introduced. From commit #440 onwards, a new testing strategy is followed, whereby the introduction of new production code (a red dot) almost always entails the immediate addition of a new unit test (a green dot). From #670 onwards, integration tests appear. Commit #670 is also interesting because it is the first indication of a "phased testing approach", signaled by the vertical yellow line, indicating that a large number of unit tests are modified. This pattern returns around commit #780 and can also been seen in the form of a large number of test additions around commit #870 and #1375.

**Growth history view.** The testing effort undertaken during Checkstyle's history can be best described as rather synchronous, as can be deduced from the curves which grow together (Figure 6). The figure confirms the initial single test code file that gradually grows and gets extensively enforced after release 2.2 (during a phase of pure testing; see annotation 1). In the period thereafter (release 2.2 and beyond), development and testing happen synchronously, with an additional effort to distribute test code over multiple classes. Increases as well as decreases in the number of files and code in production are immediately reflected in the tests most of the time, with the exception of a phase of pure testing before release 3.0 (ann. 2). This development approach is maintained until approximately halfway between release 3.1 and 3.2, where a period of pure development results in a testing time backlog (ann. 3). Thereafter, testing happens more phased until 3.5 (ann. 4). In the last period, the co-evolution is again synchronous, with a gradually decreasing time delay towards the last considered version.

In the figure, we also observe test refactorings (ann. 5).

**Coverage evolution view.** Checkstyle's coverage evolution view in Figure 7 shows a generally relatively high level of test coverage, with class coverage hovering around 80% and climbing towards 95% towards the later versions of the software. For the other levels of coverage, a similar steady increase can be seen. This trend is also confirmed by the log messages from the developers: they regularly check (and maintain or increase) the levels of coverage of their application with the help of the Clover[11] test coverage tool.

---
[7]http://svnkit.com/
[8]http://emma.sourceforge.net/
[9]http://swerl.tudelft.nl//testhistory
[10]http://checkstyle.sourceforge.net/

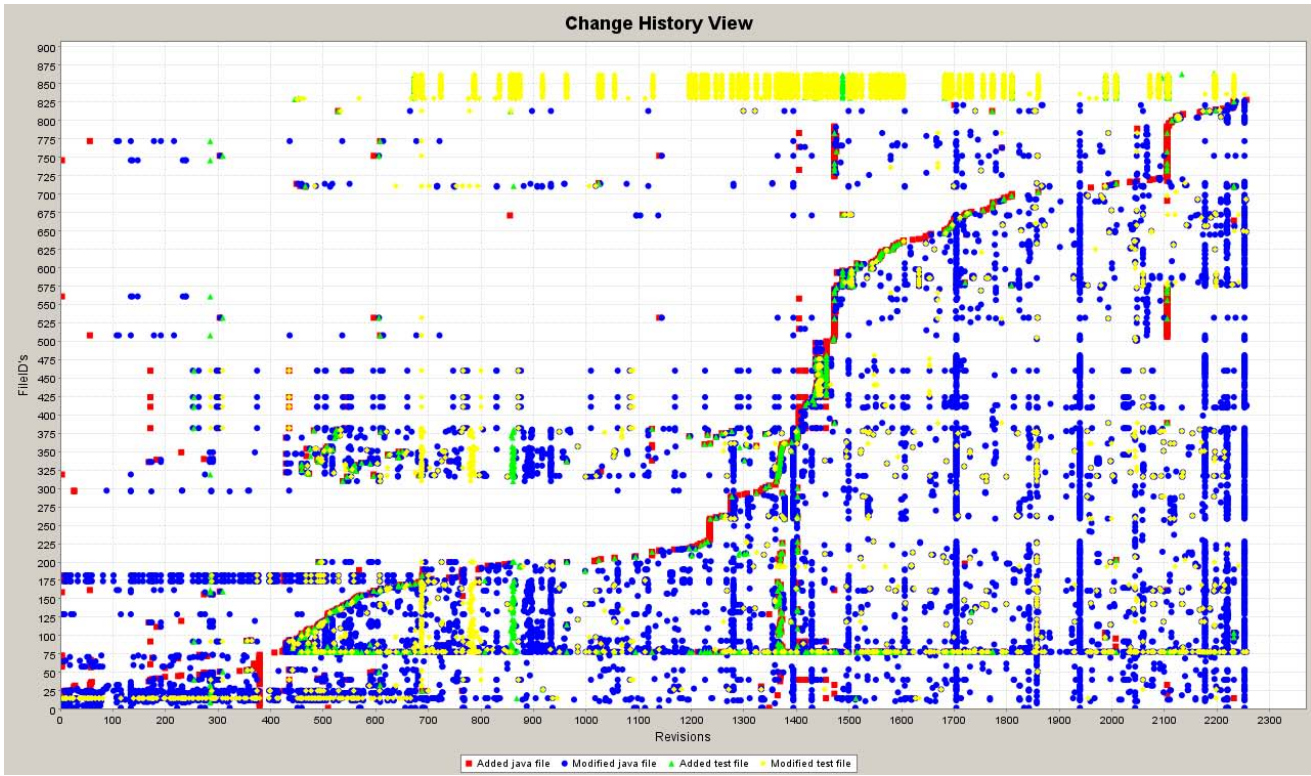

**Figure 5. Checkstyle change history view.**

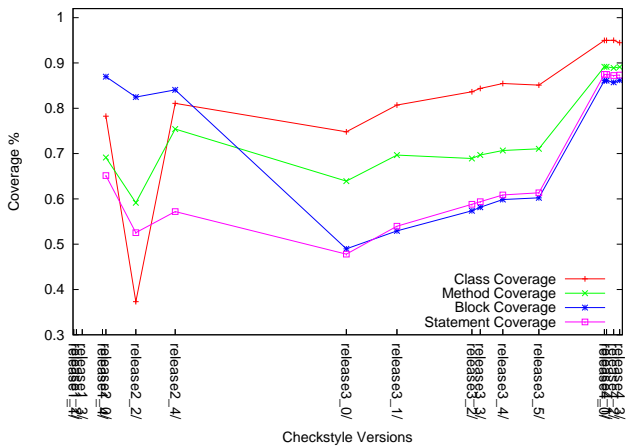

**Figure 7. Coverage Evolution View of Checkstyle.**

Around release 2.2 however, an interesting phenomenon can be witnessed: a sudden sharp decline for class, method and statement coverage, with a mild drop of block coverage. Inspection reveals that this drop is due to the introduction of a large number (39) of anonymous classes, that are not tested. These anonymous classes are relatively simple and only introduce a limited number of blocks per class, and therefor, their introduction has a limited effect on the block coverage level. Class coverage however, is more affected because the number of classes (29) has more than doubled with the 39 additional anonymous classes. Taking the inspection one step further taught us that the methods that are called within the anonymous classes *are* tested separately.

Towards the next version, all levels of coverage increase because of the removal of most of the anonymous classes.

## 5 Case 2: PMD

**Introduction.** PMD [12] is a static analysis tool that looks for potential problems in Java code, such as dead code, duplicated code, suboptimal code, etc. Its history dates back to June 2002 and, since then, 3536 subversion commits were registered (up to March 2007). Regarding the size of the project: it contains 844 classes and 56KLOC. 19 developers were involved over the course of this project.

---

[11]http://www.cenqua.com/clover/

[12]http://pmd.sourceforge.net/

**Change history view.** Figure 8 shows the behavior of the PMD developers. Of particular interest here is that a red dot is often closely followed by a green dot, meaning that the addition of production code is followed by the addition of an accompanying unit test. Also of interest are the numerous yellow vertical bars, similar to those that we observed with Checkstyle. This again supports the theory that testing is concentrated around periods with intense testing. Furthermore, we observe that in the periods in between these testing bursts, the number of changes to test code are few. When compared with Checkstyle (Figure 5), we also see that the number of integration tests is much smaller.

**Growth history view.** When considering PMD's growth evolution we observe that, in general, production and test code do not evolve synchronously: periods of pure production code development are alternated with periods of pure testing. This confirms our earlier observation from the change history view of PMD (Figure 8), where we observed that testing is mainly concentrated in short time-intervals and that in the intervals that lie in between testing bursts few test-related commits happen. An exception to this observation is the fact that the addition of new units of production code also triggers the addition of a new unit test.

**Coverage evolution view.** The test coverage at the time of releases has been slowly increasing over time. Class coverage hovers around 80%, method coverage is between 60 and 70%, and statement coverage is between 40 and 53%. This steady increase indicates that the developers do take the test coverage seriously, but at the same time, they are not reaching the same high levels of coverage that we could see with Checkstyle or JPacman.

## 6  Case 3: ArgoUML

**Introduction.** ArgoUML[13] is an open source UML modeling tool that includes support for all standard UML 1.4 diagrams. The first contributions to ArgoUML go back to the beginning of 1998, and up to December 2005, 7477 subversion commits were registered. The final release we considered for this study was built by 42 developers who wrote 1533 classes totaling 130KLOC.

**Change history view.** The change history view of ArgoUML is in line with what we saw in our previous case studies. Testing efforts are initially limited, and it is only later on that more and more tests are added. Again, we notice a significant number of yellow vertical bars, indicating a phased testing approach.

**Growth history view.** This phased testing is confirmed by the growth history view. The stepped curves for testing confirm the presence of pure testing periods, as these steps do not correspond with increases in production code (Figure 9). Besides these periods of testing, the test code is barely modified, except for the creation of test skeletons in the early history (between releases 0.10 and 0.12, see annotation 1) and periodical test refinements (ann. 2) and refactorings (ann. 3).

---

[13]http://argouml.tigris.org/

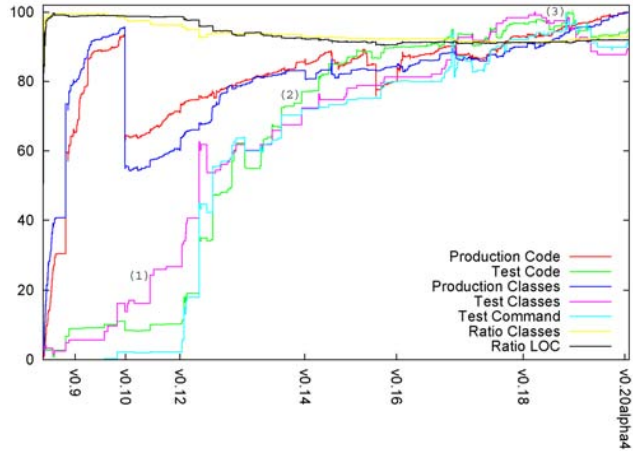

**Figure 9. Growth history view of ArgoUML.**

Note that the initial "hill" in the production code curve is due to architectural changes which are reflected in a changed layout in the versioning system, resulting in the source code residing in two locations at the same time. Later on, before release 0.10, the old layout structure and code remains get deleted.

**Coverage evolution view.** Even without this side-effect, the initial testing effort is rather low and only slowly increasing. Overall, ArgoUML has the lowest coverage of the four considered projects; it is however also the oldest and largest project. The fact that the first release of jUnit (beginning of 1998) more or less coincides with the start of the ArgoUML project might explain why the effort that went to testing was rather low in the earlier phases, as jUnit was not yet well known at that time. The last considered version of ArgoUML is characterized by a sudden drop in test coverage. This is due to the extraction of the `mdr` component, a storage backend, into a separate project. Apparently, this component was better tested than the remainder of the project, resulting in the coverage drop.

## 7  Characterizing test coverage

In the previous sections we have seen how the change history view, the growth view, and the coverage view can help to understand the testing habits in development process for three different projects. In this section, we study how the data gathered from these case studies can be combined to offer a benchmark on the relation between the fraction of test code on the one hand, and the obtained test coverage on the other, which will help us to answer question RQ5.

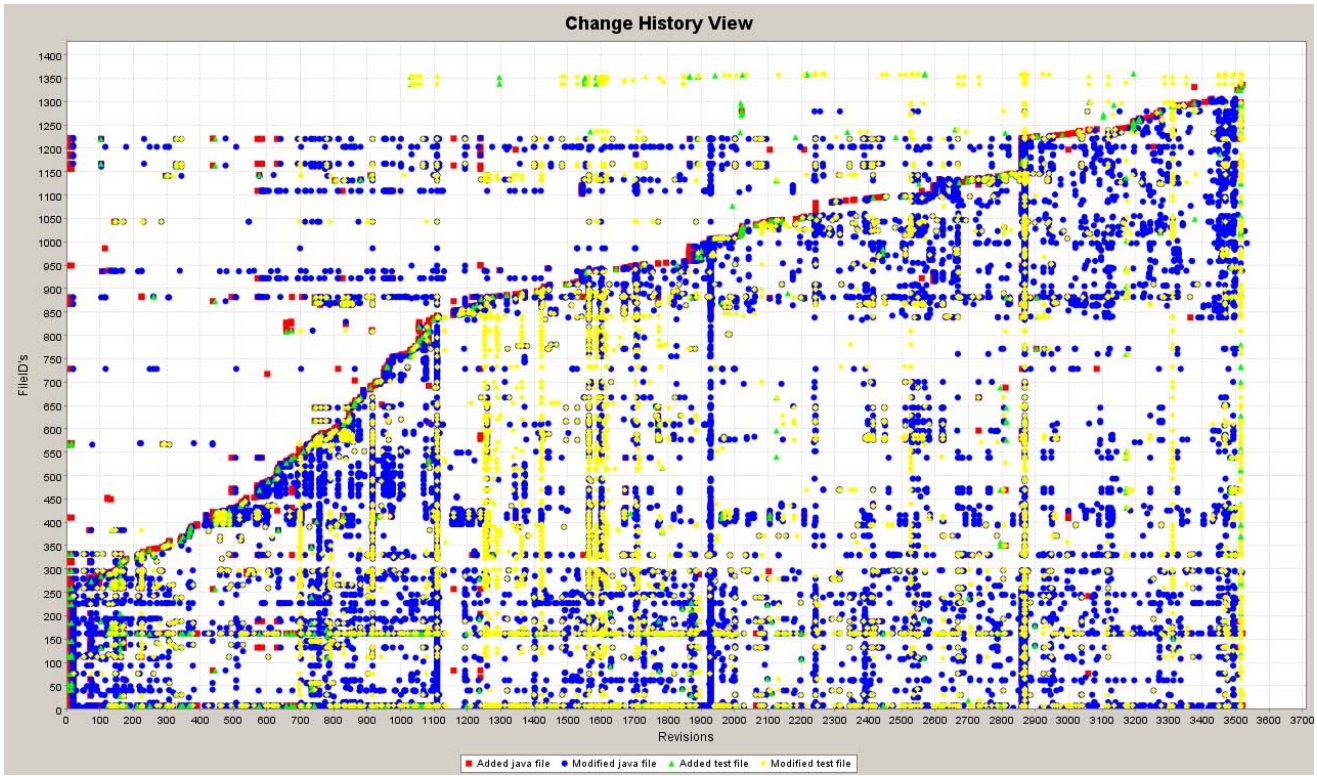

**Figure 8. Change history view of PMD.**

The contribution of test code has been reported to vary between 33% and 50% of the overall system size [10, 23]. From our data we can (i) compare the fraction of test code ($tLOCRatio = 100 - pLOCRatio$) these studies reported against the numbers we obtained; and (ii) observe whether a relation between $tLOCRatio$ and the resulting test coverage exists (RQ5).

Figure 10 presents the 51 data points obtained from the coverage measurements of the releases that we considered from the four case studies, at the coverage levels class, method, block and statement. With this data set we cover a broad spectrum, with a $tLOCRatio$ between 6.5% (Checkstyle 2.4) and 39% (JPacman 2.2.1) and test coverage percentages between 8.9% block coverage (again, ArgoUML 0.14) and 100% class coverage (all JPacman releases).

As an initial observation, we indeed notice the general trend that test coverage increases alongside test code share. To quantify the level of correlation between the variables *tLOCRatio* and test coverage, we computed Pearson's product-moment correlation coefficient ρ (Table 2). The table confirms the presence of a considerable, positive correlation ($\rho \geq 0.69$) for three of the coverage levels.

We attribute the weaker correlation for block coverage to the differences in testing approach between the projects. The use of a test coverage tool and testing efforts to increase coverage make the test code of Checkstyle very efficient (in release 2.4, 6.5% test code yields 81%/75%/84%/57% test coverage for the four considered levels). In the case of ArgoUML release 0.14 however, a similar *tLOCRatio* (about 7%) only results in 22%/14%/8.9%/10% test coverage. The developers of this system, as we have already observed in earlier work [24], apply a more integration-like kind of testing. Block coverage, being one of the finer testing levels considered here, is most impacted by such differences across projects.

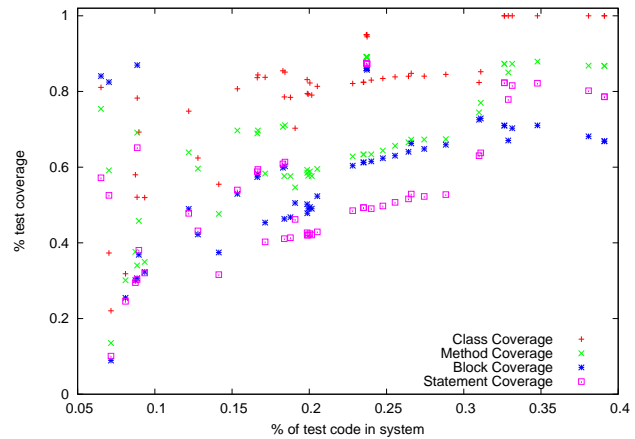

**Figure 10. Relation between Test Code share and Test Coverage.**

| Test Coverage Level | ρ |
|---|---|
| Class Coverage | 0.79 |
| Method Coverage | 0.74 |
| Block Coverage | 0.51 |
| Statement Coverage | 0.69 |

**Table 2. Correlation between Test Code share and Test Coverage.**

## 8  Discussion

We now address the research questions that we have defined in Section 1.

**RQ2** *Does co-evolution always happen synchronously or is it phased?* From both the change history view as well as the growth history view we observed visual patterns indicating the type of co-evolution a case study underwent. Specifically, in the change history view we witnessed (i) green or yellow vertical bars indicating periods of pure testing (e.g. Checkstyle, PMD) and (ii) green dots on top of red dots as indicators for the simultaneous introduction of production code with corresponding unit tests (e.g. Checkstyle). With regard to the growth history view, we saw (i) curves following each other closely denoting synchronous activities (e.g. Checkstyle), while (ii) stepwise curves point to a more phased testing approach (e.g. ArgoUML).

**RQ3** *Can an increased testing effort be witnessed right before a major release or other event in the project's lifetime?* From the case studies that we performed, we saw no evidence of a testing phase preceding a release. This is however not in line with the findings of Hindle et al. [16]. To characterize a project's behavior around release time, they partition files in the version control system into four classes: source, test, build and documentation. Before MySQL releases, increased testing and documentation effort was observed. We attribute this difference in observation to the selection of case studies. MySQL, a mature open source project that is backed up by the MySQL AB company for commercial licensing and support, has a rather strict release policy which requires severe bugs to be fixed and tested before a release can happen[14]. In contrast, our case studies entail projects that developers work on in their free time. The FAQ of ArgoUML explicitly mentions that the system is not production ready, despite more than nine years of development[15].

**RQ4** *Can we detect testing strategies, e.g. test-driven development?* From a commit perspective, test-driven development is translated as a simultaneous commit of a source file alongside its unit test. We saw evidence of test-driven development in the Checkstyle and PMD case studies, by means of green dots on top of red dots in the change history view. However, yellow dots on top of blue dots (signifying co-evolution after introduction), is not the de facto way of evolution in these projects.

**RQ1** *How can we summarize the co-evolution of test code and production code?* The combination of the three views that we introduced in this paper allowed us to observe and characterize the co-evolution of production code and test code. This claim is strengthened by the fact that we provided answers to research questions RQ2, RQ3 and RQ4.

**RQ5** *Is there a relation between test effort and test coverage?* For the four considered case studies, we computed the correlation, for every considered release, between the *tLOCRatio* and each of the four coverage levels. Using Pearson's ρ, we observed a considerable correlation. This might seem surprising as we did not take into account the following factors:

- *Kind of tests* under consideration. We took the overall coverage level into account, without making a distinction between unit tests and more integration kind of tests. For the case studies considered here, we noticed that Checkstyle has more integration tests compared to the other case studies. ArgoUML has, next to the unit test suite, a separate suite of automated GUI tests.
- The *quality focus* of the developers of the respective projects. In the change log messages of Checkstyle, developers mention the use of a coverage tool to detect opportunities for increases in test coverage. Compared to a system with a similar fraction of test code, we noticed a considerable yield in test coverage.
- The *testability* of the software system under test. Bruntink and Van Deursen observed a relation between class level metrics (especially Fan Out, Lines Of Code per Class and Response For Class) and test level metrics [4]. This means that the design of the system under test has an influence on the test effort required to reach a certain coverage criterion.

## 9  Related work

We did not find any research specifically related to the co-evolution of production code and test code. However, Hindle et al. studied the co-evolution of a number of artifacts — source, test, build and documentation — of MySQL [16]. Work related to each of the individual views is captured in the motivational sections.

## 10  Conclusion & future work

In this paper we observed the co-evolution between production code and test code. In this context, we made the following contributions:

---

[14] http://dev.mysql.com/doc/refman/5.0/en/release-philosophy.html (accessed April 11, 2007)

[15] http://argouml.tigris.org/faqs/users.html (accessed April 11, 2007)

1. We introduced three views: (i) the change history view, (ii) the growth history view and (iii) the coverage evolution view. We combined them to study how test code evolves over time.
2. According to our three open source case studies, testing is mainly done in phases.

Indeed, all our cases show time intervals of pure testing, intertwined with periods of pure development. Synchronous co-evolution happens seldomly. We also did not observe testing phases right before a release. Evidence of test-driven development was found, as we saw numerous unit tests being introduced alongside their corresponding production code. Using case studies with different levels of test coverage, we observed a large variation in the fraction of test code needed to reach a certain level of test coverage.

As for future work, it is our aim to extend this research to industrial software projects, as the results might differ greatly in a context where imposed testing standards are in place. Another step we want to take is get a deeper insight into the factors that influence the relationship between the fraction of test code and the level of test coverage.

**Acknowledgments** This work has been sponsored by (i) the Eureka ∑ 2023 Programme; under grants of the ITEA project if04032 (*SERIOUS*), (ii) the NWO Jacquard *Reconstructor* project, and (iii) the Interuniversity Attraction Poles Programme - Belgian State – Belgian Science Policy, project *MoVES*.

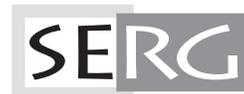